# The Fidelity and Feedback Traps: The Case for Health Digital Twins as Modular Evolving Causal Systems

Authors: Nikki L. B. Freeman[1,2,3], Yating Zou[3,4], Kyungbok Lee[3,4], Chenyao Yu[5], Didong Li[3,4], and Michael R. Kosorok[3,4,6]

1. Department of Biostatistics and Bioinformatics, Duke University, Durham, NC
2. Duke Clinical Research Institute, Durham, NC
3. UNC Gillings Center for AI and Public Health, Chapel Hill, NC
4. Department of Biostatistics, University of North Carolina at Chapel Hill, Chapel Hill, NC
5. Department of Statistical Science, Duke University, Durham, NC
6. Department of Statistics and Operations Research, University of North Carolina at Chapel Hill, Chapel Hill, NC

Corresponding author:
Nikki Freeman
Duke Clinical Research Institute
300 W. Morgan St.
Durham, NC 27701
nikki.freeman@duke.edu

**Abstract:** Digital twins for health may be used to compare treatments, project patient trajectories, and support clinical decisions. While related to mechanical digital twins, those initially developed for engineering applications, replicating the mechanical digital twin architecture and goals may fail in health for two reasons. The fidelity trap is the belief that an accurate model can answer what-if questions by virtue of its accuracy. Prediction and counterfactual reasoning are different tasks, and a twin that can fit past trajectories well may miss the mark when ranking treatments. The feedback trap arises when the twin updates on data its own recommendations helped generate. Refitting in this way can recover a biased relationship and grow more confident even as data grows thinner. We contend that health digital twins should be conceived as causally valid, modular, and evolving systems. Modularity isolates the data and models needed for interventional recommendations, causal validity supports such claims, and governed evolution updates the twin while accounting for how its recommendations reshape the data. We conclude that the standard for a health twin should be how well it supports decisions in the world it helps create, not how faithfully it reproduces the world it observes.

**Key words:** Digital twins, causal inference, clinical decision support, artificial intelligence, precision medicine

# 1. Introduction

A digital twin (DT) is a computational model of a physical counterpart.[1,2] Unlike a conventional simulation, the twin evolves with its counterpart, assimilating new observations as they arrive, and aims to support counterfactual reasoning about what would happen under interventions not yet taken.[3,4] Rooted in aerospace engineering, DTs have spread to manufacturing, energy, urban planning, and, increasingly, public health as computer simulation has grown cheaper, sensors have become ubiquitous, and data-integration tooling has matured.[1,2,5–17]

In health, the DT physical entity can be many things—an individual, a population, a biological process—and the twin can be designed to simulate treatment strategies, predict health trajectories, and support decision-making.[8,18,19] In practice, these targets and scales often do not separate cleanly: population-level data may inform patient-level decision support, while individual decisions and responses can reshape the data available for future updating. Accordingly, the data feeding a DT are often multi-scale and multimodal, and modeling strategies range from mechanistic models grounded in physiology to data-driven models learned from large observational corpora, or may use hybrid or systems-level combinations of the two.[10,12,20–23] Health applications of DT span cardiac dynamics, oncology dose selection, and sepsis management, among others.[21,24–26]

The systems surrounding a health DT and driving its evolution differ from the engineering setting in several ways. We focus on systems whose data are generated by a complex, possibly non-stationary process shaped by the very decisions the twin is designed to support. Most importantly, there is no known physical law to anchor the DT. The data-generating process is unknown and must be estimated. The process drifts over time, its inputs are partly latent, and no law guarantees that predictions beyond the observed regime are correct.[27,28] Moreover, the counterpart has agency. Patients and clinicians anticipate and respond to the twin's recommendations. The data used to update the twin are shaped by its own outputs, both mechanically and behaviorally. Additionally, a mechanical system can be reset and rerun under the same conditions, while this is impractical and often impossible for health decision-making.

Despite these differences, many existing health digital twins inherit the engineering architecture, tuned for predictive accuracy and representational fidelity.[3,8,19,29] That is, they are optimized for how closely the virtual system reproduces observed trajectories of the physical entity. This objective works when physical laws provide strong structural constraints, but in health, it falls short of the purposes to which digital twins are now being applied. Two traps explain why.

The first, the **fidelity trap**, concerns the belief that a sufficiently accurate model will, by virtue of its accuracy, support reliable what-if counterfactual reasoning. In health, a model can reproduce observed patient trajectories with high accuracy yet remain unable to say which treatment a patient should receive next.[30] This is because prediction and counterfactual estimation are distinct inferential tasks, each resting on its own justification.[31,32]

The second is the **feedback trap**. Engineering twins are closed-loop yet safe, because a physical, observed law keeps decision-influenced data honest.[4,33] Health removes that protection twice over: (1) the twin's recommendations alter which treatments are administered, which states are measured, and which alternatives remain represented in the data, and (2) the agents being modeled respond to those recommendations in turn. The observations the twin later learns from are endogenous to its own outputs, violating the identification assumptions that justified its counterfactual inferences in the first place.[34,35]

We argue that digital twins for health must be reconceived as causally valid, modular, and evolving systems, and that these properties are what keep a twin out of the two traps. Causal validity and modularity together answer the fidelity trap: modularity isolates the components that require interventional justification from those that do not, and causal validity supplies that justification explicitly rather than inferring it from predictive fit. Principled evolution answers the feedback trap by updating the twin in a way that accounts for how its own recommendations and the responses they provoke shape the data it observes.

This paper, developed from Michael Kosorok's plenary presentation at the 2025 American Causal Inference Conference session on the Frontiers of Causal Inference, proceeds as follows. Section 2 introduces the fidelity and feedback traps using dynamic insulin dosing as a running example. Section 3 develops the proposed causally valid, modular, and evolving framework and explains why adjacent approaches address only parts of the problem. Section 4 discusses implications for researchers, clinicians, system designers, and regulators, and how the framework extends beyond the running example. Section 5 contains the discussion.

## 2. The fidelity and feedback traps

In this section, we further define the fidelity and feedback traps and argue why they are detrimental to health DTs. To ground our discussion, we use the example of dynamic insulin dosing from continuous glucose monitoring (CGM) data, where the physical counterpart is a patient-level glucose-management process. The relevant DT task is not merely to forecast the next CGM value, but to continuously support dosing decisions by projecting future glucose trajectories or safety-relevant summaries, such as time in range and hypoglycemia risk, under candidate insulin choices.

### 2.1 The fidelity trap

The fidelity trap is the assumption that a sufficiently accurate model supports reliable what-if reasoning by virtue of its accuracy. Predictive accuracy tells us a model fits observed data, not what would have happened under a treatment the patient did not receive. To inform actions, DT outputs must be tied to counterfactual quantities that are identified from observed data under explicit assumptions.[31] Moreover, we often need the DT to support multiple counterfactual claims, each with its own assumptions and supporting data.

In glucose management, larger insulin doses may be followed by higher glucose because they are used when glucose is already high, rising, or otherwise difficult to control.[36] This is not evidence that insulin worsens glucose control. Instead, it reflects how historical dosing is tied to a patient's state. Adding richer history, such as recent glucose trend, prior insulin, meals, and activity, can help address this problem, but it does not, by itself, turn observed-path prediction into a valid comparison of alternative dosing choices.[37] Consequently, a predictive twin may reproduce the historically chosen dosing path in similar situations rather than support a causal comparison of the choices it was asked to weigh. For example, if the twin is used to compare no correction, a smaller correction, and the usual dose, accurate prediction of the historical path does not establish whether either alternative dose would lead to a better outcome.

## 2.2 The feedback trap

The feedback trap arises when the twin is updated using data that its own deployment helped generate.[34] Updating can seem unambiguously beneficial, and for a mechanical twin, this can be true. Physical laws govern the physical entity, and the law itself does not change when actions are taken under the twin's influence, so refitting on decision-influenced data recovers the same mechanistic system. Decision-making in health has no such law. The functional form is itself unknown and estimated from data, and because the actions in that data were selected in ways tied to patient state and history, refitting can recover a biased relationship for the choices that were made while saying little about the alternatives the twin must compare.[35] Further, the observed data depend on how agents in the physical system respond to recommendations and what the twin prompts us to look at, since what, when, and how often to measure are themselves decisions the twin shapes.[34]

We illustrate two examples of failing to account for feedback. First, the twin can confound the very outcomes used to judge it. In glucose management, if a twin marks a patient group as low risk and recommends less frequent monitoring, fewer glucose trajectories for that group are recorded. The next refit reads this thinner record as evidence of stability and lowers monitoring further. Methods that explore when uncertain cannot help here,[38,39] because the censored data make the twin look more certain, not less. Second, the recommendation issued by the twin may not be the treatment that is actually delivered. A patient may delay, reduce, or skip a recommended dose because of meal uncertainty, planned activity, symptoms, or fear of hypoglycemia. A clinician may override the recommendation because of clinical judgment or competing priorities.

This gap is created by deployment, and it can shift as patients and clinicians adapt to a twin they have come to anticipate.

## 3. Causally valid, modular, evolving systems

The two traps point to two design requirements. To avoid the fidelity trap, each recommendation must be tied to a causal claim, the identification assumptions it rests on, the data that support it, and its quantified uncertainty. To avoid the feedback trap, the twin must monitor how its own recommendations change treatment, measurement, patient and clinician behavior, and the support available for future comparisons. The proposed framework separates the inferential roles needed to keep each decision-support output tied to a traceable causal claim. These roles span identifying the decision context, comparing interventions, managing uncertainty, recommending or falling back, and governing post-deployment. How such a system addresses the fidelity and feedback traps is described in the following and visualized in Figure 1.

### 3.1 Causally valid and modular: answering the fidelity trap

We conceptualize a health digital twin as a network of models and data. The network can be small, a single model fit to a single dataset, or it can scale, with many model components, each with its own data and interconnected so that the outputs of one model are inputs to others. This is pragmatic: new data sources can be woven in, components updated as methods improve, or pieces may be dropped when a twin is transported to a setting where data are unavailable.

Beyond flexibility, a twin may need to make several causal claims, possibly on different scales, for different populations, in different contexts and settings, and under different interventions and policies. The same underlying twin may be queried to support a short-horizon comparison of candidate actions, a longer-horizon comparison of strategies, or a population-level evaluation of an allocation rule. Whether a given causal claim is valid depends on the data, assumptions, and modeling choices it draws on.[31]

Because of this, the twin must be organized so that the data, assumptions, uncertainty, and other components can be flexibly grouped and traced. This is what we mean by modularity. This structure also shows what additional data, assumptions, or design changes would be needed to move a partially identified or unidentified claim toward a more informative causal statement.[40] The state representation is one of the things grouped this way. In health, the states that drive an action's effect must be discovered from data and revised as understanding improves. Grouping by claim lets these states be found locally, for the decision and population the claim concerns, rather than fixed once for the whole twin. Personalization follows from this, allowing the model, and hence the physical system, to use different parts of the available information to best inform different types of decisions.

Recall the glucose management confounding by indication example from Section 2.1. Viewing the twin in a modular way lets us separate the causal estimate from the rest of the twin, so the confounding is isolated in the components where the dose effect is claimed. Once located, appropriate modeling choices can be made to address it. This separation prevents future glucose prediction, candidate-dose comparison, uncertainty assessment, and decision support from being collapsed into a single predictive task. In this way, the model aims to provide decision-support output anchored to an explicit comparison among candidate doses, with the uncertainty attached to that comparison.

## 3.2 Evolving and living: answering the feedback trap

The modular architecture provides a causally valid foundation at any fixed point in time. To be a living system, the twin must also revise itself as new observations arrive. The difficulty is that once the twin is deployed, those observations are no longer generated independently of the twin.[34] They are partly shaped by the recommendations the twin has issued and by the ways patients and clinicians respond to them.

Evolution must therefore be governed, and this governance operates at two levels. At the scale of a single component, this means re-estimating, recalibrating uncertainty estimates, and, when warranted, revising the identification strategy that the component depends on. At the scale of the whole twin, it means deciding when accumulated discrepancy calls for re-estimation, when a structural shift calls for recomposing components, and when a claim must be suspended pending more data, much like an adaptive stopping rule in a sequential trial.[41–43] This governance layer can itself be a model in the network, one that monitors the other components and makes these calls, with human oversight for the consequential ones.

Governance decides when the twin updates and when it holds back. When the data continue to support the relevant intervention comparison, ordinary updates may be enough. When uncertainty drifts, treatment options lose support, or the gap between the recommendation and delivered care widens, the twin should trigger governance actions such as recalibration, re-estimation, claim narrowing, deferral, or fallback to a conservative rule.[28,44] Because the counterpart is an agent, the updated rule must also account for how recommendations change behavior, so the twin adapts to the realized distribution rather than passively projecting the past.

Uncertainty is the harder thing to govern, because it does not sit in any single component. It has to be carried across every module a claim depends on; an error in one surfaces in the final decision. In glucose management, if the twin underestimates recent insulin on board, an aggressive correction dose can look safer than it is. The comparison that ranks doses inherits the bias, and the dose is ranked too highly. An uncertainty estimate built on already-biased projections may not reveal the problem. Uncertainty must also be tracked across modules and decision timescales on which the twin acts and re-measures.[29] The timing itself, when to act, when to re-measure, and when to re-decide, is an output of the twin that carries its own uncertainty. Modularity does not remove these risks, but it makes the interfaces visible, so uncertainty, drift, and failure can be monitored where they arise.

Governed evolution is what answers the feedback trap. It treats deployment itself as part of the data-generating process. The twin does not simply learn from new data; it learns from response patterns that may themselves have been altered by its own recommendations.[34] Maintaining joint causal validity across such an evolving modular network remains a foundational problem, and the digital twin setting makes that problem concrete.

### 3.3 Relation to adjacent approaches

One might ask whether the framework outlined above is already supplied by existing approaches. We have contended that a digital twin for health should be causally valid, modularly composed, and designed for principled evolution in response to feedback. Established fields provide important tools for these goals, but none supplies all three simultaneously within a deployed, evolving health DT. Model predictive control selects actions by simulating each candidate forward in time using a dynamics model and comparing the resulting trajectories.[45] In health decision-making, the dynamics are usually unknown and must be estimated from confounded data, a task for which model predictive control does not by itself provide a strategy. Causal inference enables the estimation of effects under explicit assumptions, but in health DTs, the data-generating process may evolve and be shaped by the twin's own outputs.[31,34] The estimand and dataset are not fixed. Reinforcement learning provides a framework for learning to act from sequential data, but often relies on exploration to generate the data it needs.[46] A health twin cannot explore freely because unrestricted exploration may be unsafe, and the counterpart is an agent that responds. When confined to logged data, it inherits the confounding and support limits of the policy that produced those data.[47] Trial designs, such as SMARTs, provide structure for generating personalized treatment strategies, but they rely on prospective randomization to support valid comparisons and to ensure overlap by design.[48] A deployed clinical twin generally does not run a randomization protocol. Thus, each field leaves a gap when applied to a deployed health twin. The way forward likely draws on all of them, but what remains missing is a framework for maintaining causal validity across linked modules as the twin updates under feedback.

## 4. Why this matters and how it scales

Reconceiving digital twins as causally valid, modular, evolving systems has practical consequences for health decision-making stakeholders. For researchers, it changes what counts as a valid benchmark. A twin can match observed glucose trajectories closely and still rank doses by confounded association, so predictive accuracy on held-out trajectories cannot certify a twin whose purpose is counterfactual reasoning. A discriminating benchmark must score a held-out interventional contrast, ideally anchored to experimental or trial ground truth, so that a predictively excellent but causally wrong twin fails it. Validating an evolving twin means tracking its identification assumptions over time, not just its predictive fit. Partial identification is an acceptable and often necessary outcome, yet we still need the bounds to be sufficiently narrow to be useful.

For clinicians and system designers, it specifies what a recommendation should include: the interventional contrast it rests on, the identification assumptions invoked (and which are assumed versus supported), a calibrated interval for the projected outcome, and the explicit condition under which the twin defers to a conservative fallback. The twin might defer, for instance, when the interval crosses a hypoglycemia threshold or grows too wide over the planning horizon. It also specifies what a deployed twin should monitor: a recommendation-to-realization interface that tracks the issued recommendation, the delivered treatment, and the contexts in which the two diverge. Used as a monitoring module, this interface can help users recognize repeated delays, modifications, overrides, or decisions not to follow the recommendation, and can flag settings in which post-deployment data may no longer support the same intervention comparisons the twin was designed to make. These patterns can then be returned to researchers and system designers as part of post-deployment review.

For regulators, an evolving causal twin raises questions that differ from those posed by a predictive device. A submission or label should state the causal claims and the assumptions behind each, name the estimand and its identification conditions, and specify how deployment-induced distribution shift will be monitored, with explicit triggers at which claims are suspended, narrowed, or re-estimated.[27,49,50] One such trigger is a measured collapse in overlap for a candidate action.[28]

The glucose example is one instance of a general pattern. We have developed it at the scale of a single patient, but the same structure holds as the physical counterpart grows, from an individual to a clinic, a hospital, or a community. It carries over to settings such as mechanical ventilation, oncology dose titration, and treatment sequencing in chronic psychiatric conditions. The framework is architectural rather than disease-specific. The modules, estimation strategies, and identification assumptions differ by setting, while the requirements of causal validity, modular composition, and principled evolution under feedback are shared.

## 5. Discussion

The central distinction we have drawn is between a twin governed by a known physical law and one that is not. A mechanical twin can be optimized for fidelity because the law constraining its counterpart holds regardless of the twin's actions, so refitting on decision-influenced data recovers the same system. In that way, health decision-making is not comparable; predictive accuracy no longer supports counterfactual reasoning, and updating on newly observed data no longer preserves validity since the twin’s own recommendations shape the data. We call these two implications the fidelity and feedback traps, respectively. We have therefore argued that a health digital twin should be organized around decision-support outputs whose causal claims are explicit, traceable to the modules that carry them, and governed as deployment reshapes the data-generating process.

What is new here is not the constituent problems but their composition. Confounding, distribution shift, and feedback are each well understood in isolation; the proposed architecture presupposes that the identification of causal targets is carried out module by module, under stated assumptions. Its contribution is to make those assumptions traceable, to confine a failure to the module where it occurs, and to fall back to partial identification where point identification is indefensible.

Health digital twins are already being built and deployed. The danger is that they succeed on the wrong terms, reproducing observed trajectories closely enough to be trusted while ranking interventions whose outcomes are confounded by association, and growing more confident as the data they shape grow thinner. Treating a twin as an evolving causal system forces each recommendation to name the comparison it rests on and the condition under which it should defer. Maintaining that validity jointly, across a network that keeps changing under its own influence, remains an open problem. But it is the problem that health digital twins must be designed to solve: The standard for a health digital twin, then, should not be how faithfully it reproduces the world it observes, but how responsibly it supports decisions in a world it helps to create.

**Author contributions:** NLBF and MRK conceived the paper. NLBF, YZ, KL, CY, and DL drafted the manuscript. YZ, CY, and NLBF developed the figure. NLBF and YZ revised the manuscript. MRK critically reviewed the manuscript for important intellectual content. All authors read and approved the final manuscript. NLBF is the guarantor.

**Acknowledgments**: KL was supported by the National Center for Advancing Translational Sciences (NCATS), National Institutes of Health, through Grant Award Number UM1TR004406. The last author was funded in part by the Gillings Center for AI and Public Health at the University of North Carolina at Chapel Hill. NLBF, YZ, CY, and DL declare no relevant funding.

**Competing interests**: The authors declare no financial or non-financial competing interests.

**Data availability**: Data sharing is not applicable to this article as no datasets were generated or analyzed during the current study.

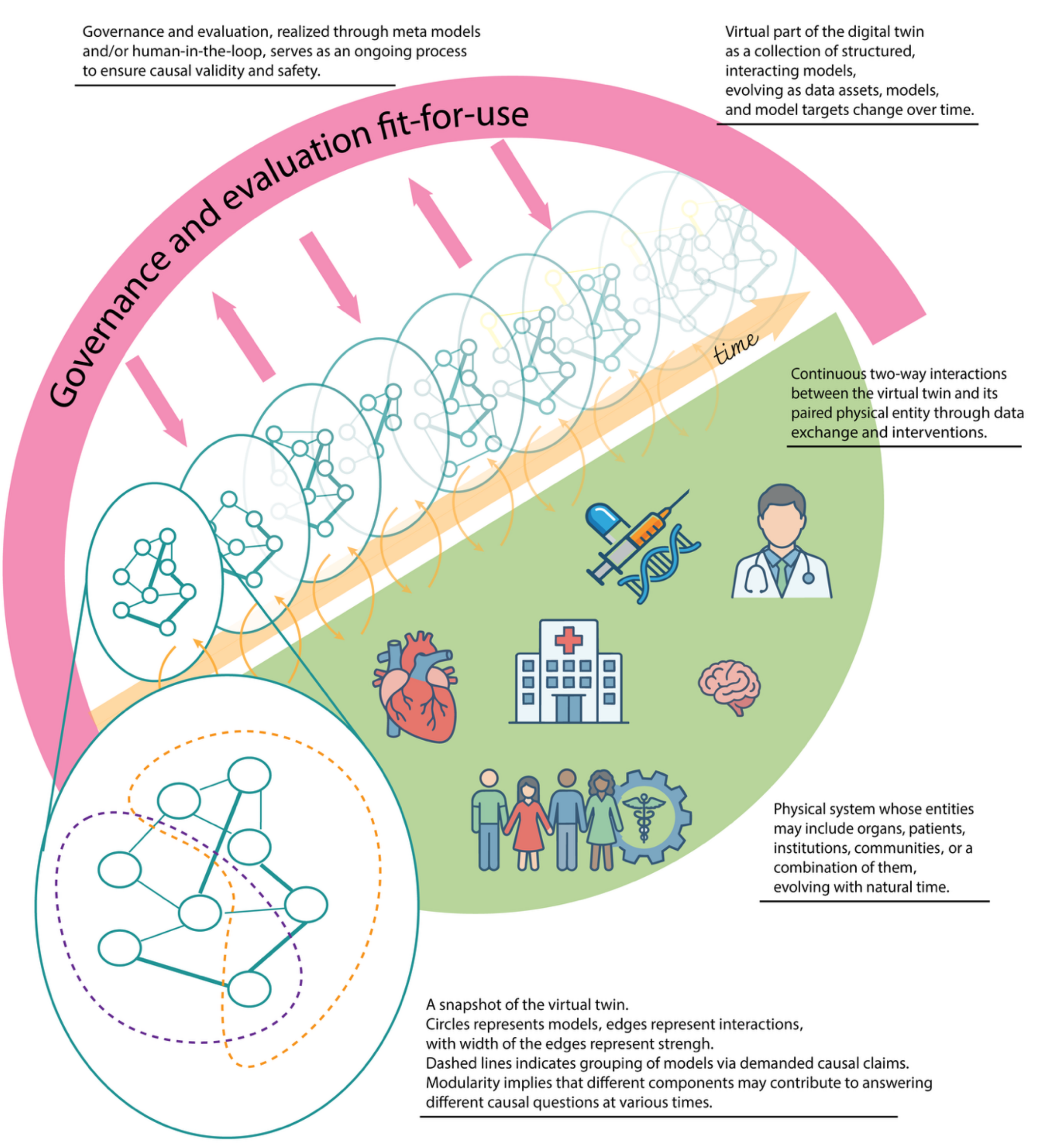

**Figure 1. Conceptual schematic of a causally valid, modular, and evolving health digital twin.** A virtual twin composed of interacting models (teal; circles are models, edges are interactions, and edge width denotes interaction strength) is continuously and bidirectionally coupled (orange) to a physical system (green) whose entities may include organs, patients, institutions, communities, or combinations of them, each evolving over time. The dashed groupings in the enlarged snapshot (lower left) depict modularity: subsets of models are grouped to answer specific causal claims, and components can be recombined to address different causal questions at different times. As data assets,

models, and model targets change, the virtual twin advances as a sequence of such snapshots along the time axis. Overarching this evolution, a governance and evaluation layer (pink), realized through meta-models and/or human-in-the-loop oversight, continuously works to keep the twin fit-for-use by maintaining causal validity and safety.